\documentclass{sig-alt-release2}

\begin{document}
%
\conferenceinfo{SIGIR'09,} {July 19--23, 2009, Boston, Massachusetts, USA.} 
\CopyrightYear{2009}
\crdata{978-1-60558-483-6/09/07} 

\title{K-tree: Large Scale Document Clustering}
%
%
%
%
%

\numberofauthors{2} 
%

%
%

\author{
\alignauthor
Christopher M De Vries\\
       \affaddr{Faculty of Science and Technology}\\
       \affaddr{Queensland University of Technology}\\
       \affaddr{Brisbane, Australia}\\
       \email{chris@de-vries.id.au}
\alignauthor
Shlomo Geva\\
       \affaddr{Faculty of Science and Technology}\\
       \affaddr{Queensland University of Technology}\\
       \affaddr{Brisbane, Australia}\\
       \email{s.geva@qut.edu.au}
}




\maketitle
\begin{abstract}
We introduce K-tree in an information retrieval context. It is an efficient approximation of the k-means clustering algorithm. Unlike k-means it forms a hierarchy of clusters. It has been extended to address issues with sparse representations. We compare performance and quality to CLUTO using document collections. The K-tree has a low time complexity that is suitable for large document collections. This tree structure allows for efficient disk based implementations where space requirements exceed that of main memory.
\end{abstract}

%

\section{K-tree}
The K-tree is a height balanced cluster tree. A K-tree contains nodes of order $m$, restricting the maximum number of vectors in any node. There are two types of nodes. Leaf nodes contain 1 to $m$ vectors. Internal nodes contain 1 to $m$ (vector, child node) pairs. A nearest neighbour search tree is built bottom-up by splitting full nodes with k-means. Once the tree increases in depth it forms a hierarchy of ``clusters of clusters'' until the above leaf level is reached. The above leaf level contains cluster centres and pointers to the leaves containing vectors inserted into the tree. For more information see Geva \cite{Geva2000} and the K-tree Project Page \cite{ktreeweb}.

The algorithm shares many similarities with BIRCH \cite{Zhang1997} as both are inspired by the B$^+$-tree data structure. However, BIRCH does not keep the inserted vectors in the tree. As a result, it can not be used for a nearest neighbour search tree. This makes precise removal of vectors from the tree impossible.

The K-tree algorithm was initially designed to work with dense vectors. This causes space and efficiency problems when dealing with sparse document vectors. We will highlight the issue using the INEX 2008 XML Mining collection \cite{Denoyer2006}. It contains 114,366 documents and 206,868 stopped and stemmed terms. TF-IDF culling is performed by ranking terms. A rank is calculated by summing all weights for each term. The 8000 terms with the highest rank are selected. This reduced matrix contains 10,229,913 non-zero entries. A document term matrix using a 4 byte float requires 3.4 GB of storage. A sparsely encoded matrix with 2 bytes for the term index and 4 bytes for the term weighting requires 58.54 MB of storage. It is expected that sparse representation will also improve performance of a disk based implementation. More data will fit into cache and disk read times will be reduced.

Unintuitively, a sparse representation performs worse with K-tree. The root of the K-tree contains means of the whole collection. Therefore, the vectors contain all terms in the collection. This makes them dense. When building a K-tree, most of the time is spent near the root of tree performing nearest neighbour searches. This explains why the sparse representation is slower. The most frequently accessed vectors are not sparse at all.

\section{Medoid K-tree}
We propose an extension to K-tree where all cluster centres are document exemplars. This is inspired by the k-medoids algorithm \cite{Kaufman1987}. Exemplar documents are selected by choosing the nearest documents to the cluster centres produced by k-means clustering. Internal nodes now contain pointers to document vectors. This reduces memory requirements because no new memory is allocated for cluster centres. Run time performance is increased because all vectors are sparse. The trade-off for this increased performance is a drop in cluster quality. This is expected because we are throwing away information. Cluster centres in the tree are no longer weighted and are not updated upon insertion.

\section{Experimental Setup}
Performance and quality is compared between CLUTO \cite{Karypis2002} and K-tree. The k-means and repeated bisecting k-means algorithms were chosen from CLUTO. The medoid K-tree was also used to select 10\% of the corpus for sampling. This sample was used to construct a K-tree. The resulting K-tree was used to perform a nearest neighbour search and produce a clustering solution. In all cases the K-tree order was adjusted to alter the number of clusters at the leaf level. CLUTO was then run to match the number of clusters produced. All algorithms were single threaded.

The INEX 2008 XML Mining corpus and a selection of the RCV1 corpus used by \cite{Song2008} were clustered. The INEX collection consists of 114,366 documents from Wikipedia with 15 labels.  The subset of the RCV1 corpus consists of 193,844 documents with 103 industry labels. Both corpora were restricted to the most significant 8000 terms. This is required to fit the data in memory when using K-tree with a dense representation.

Micro averaged purity and entropy are compared. Micro averaging weights the score of a cluster by its size. Purity and entropy are calculated by comparing the clustering solution to the labels provided. Run-times are also compared.

\section{Experimental Results}
\begin{figure}[!h]
\centerline{
\includegraphics[scale=0.45]{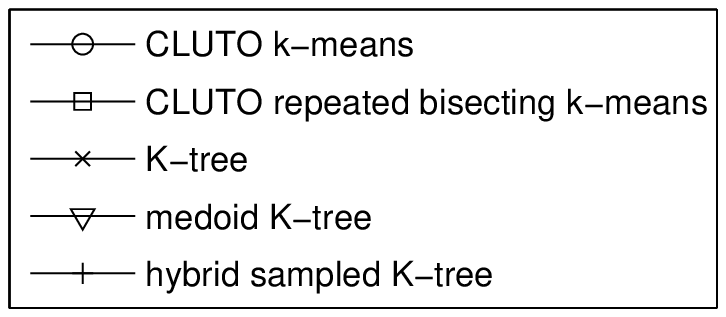}
}
\end{figure}

\vspace{-20px}

\begin{figure}[!h]
\begin{minipage}[b]{0.495\linewidth}
\centering
\includegraphics[scale=0.185]{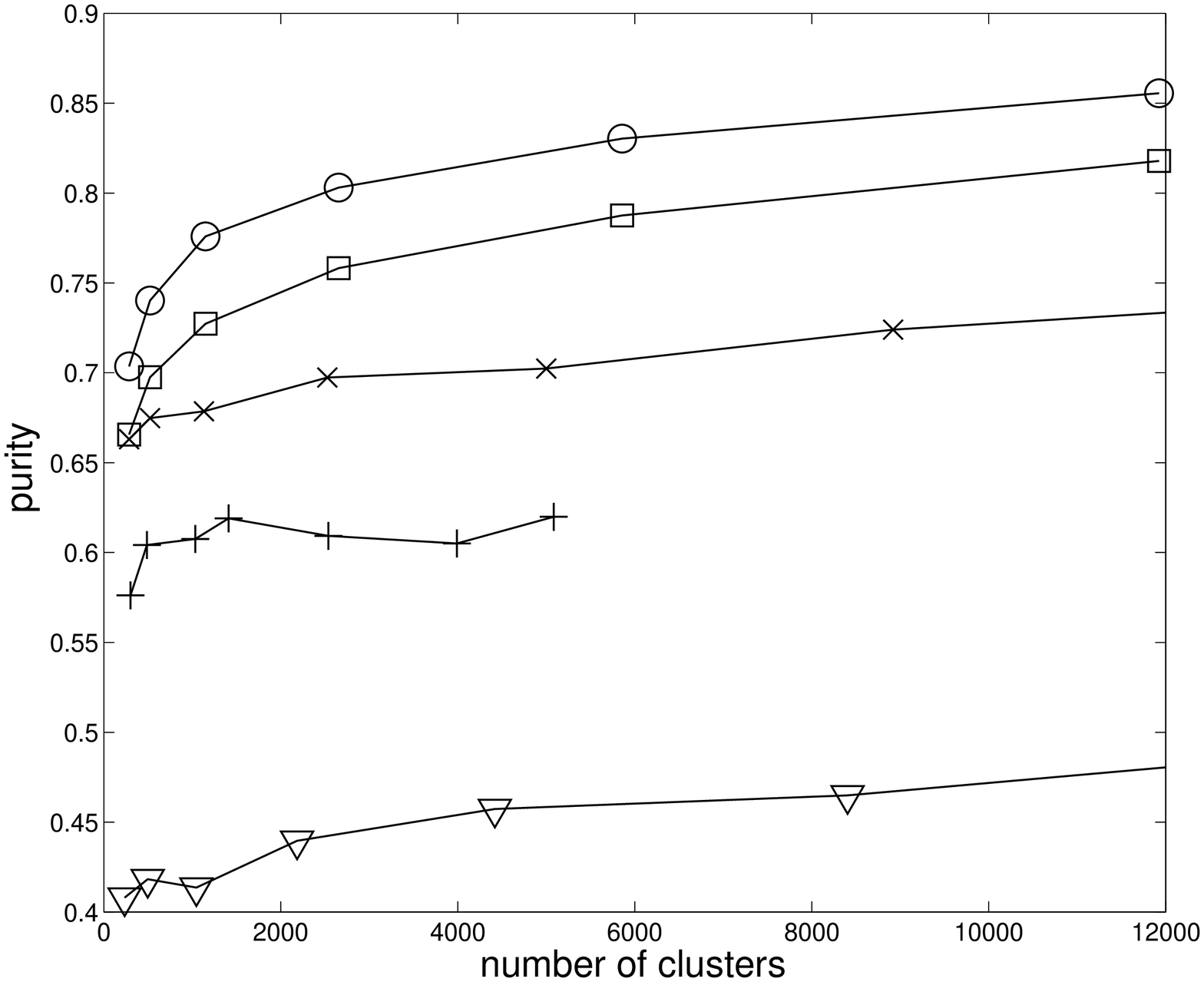}
\includegraphics[scale=0.185]{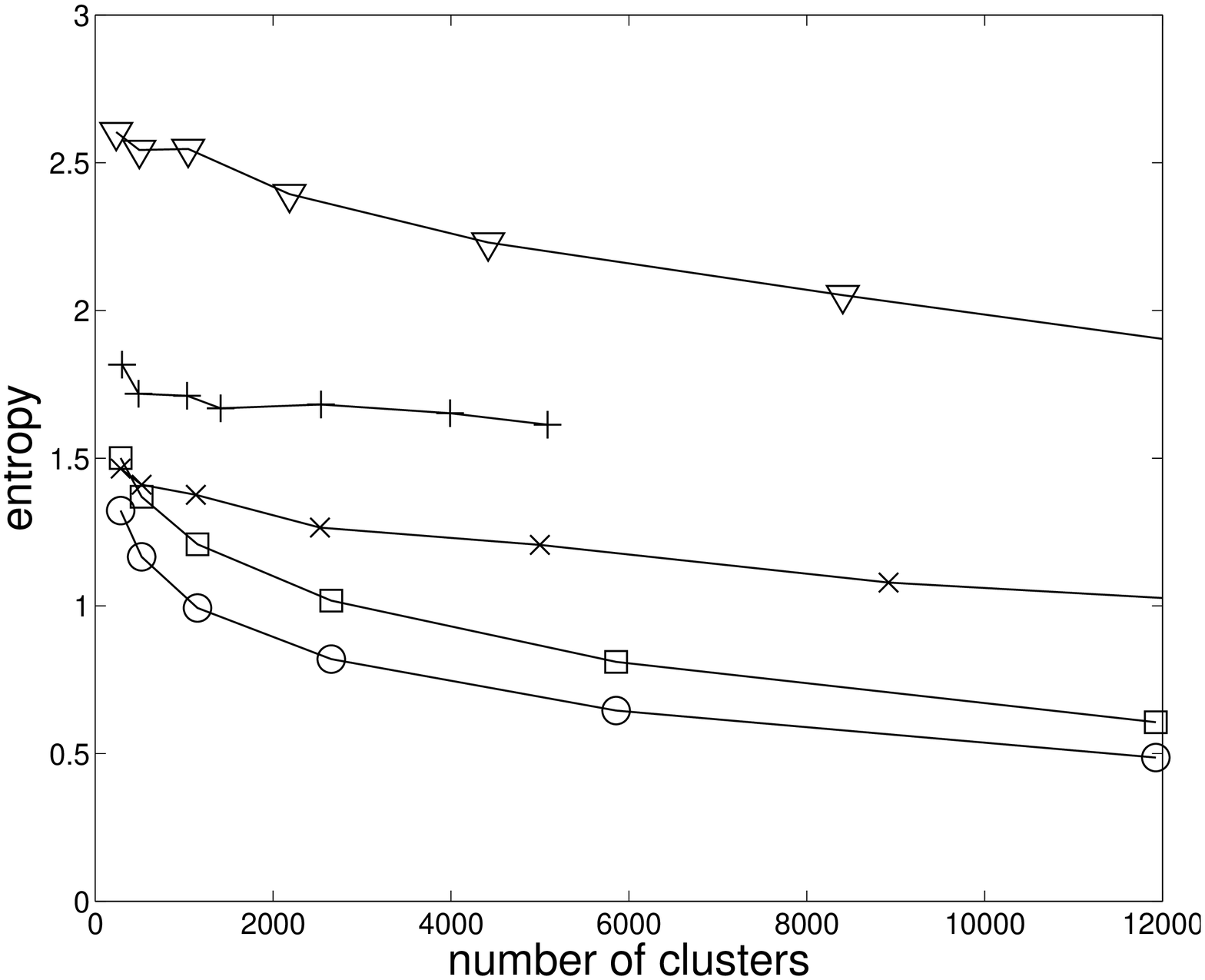}
\includegraphics[scale=0.185]{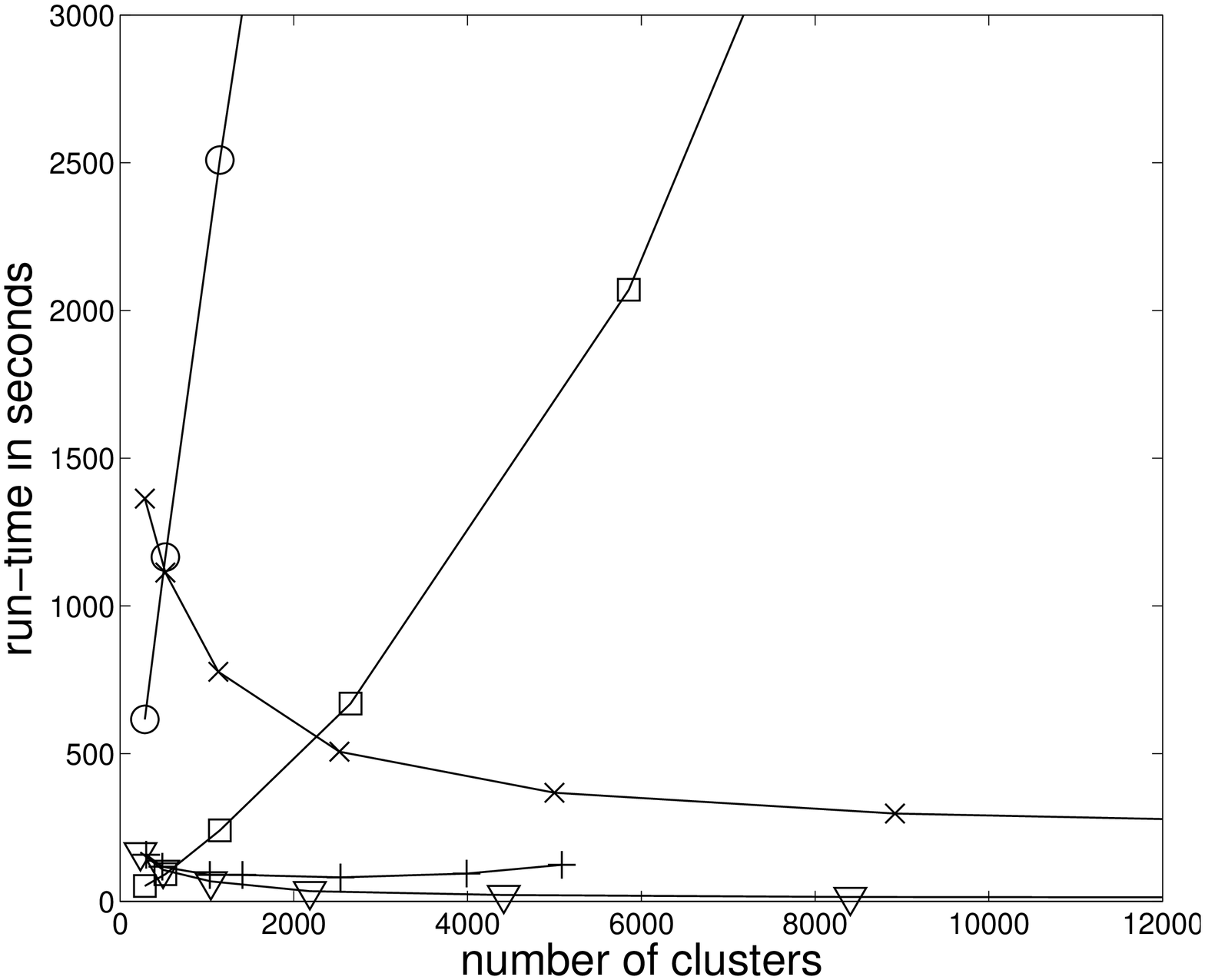}
\caption{INEX 2008}
\label{fig:inex}
\end{minipage}
\begin{minipage}[b]{0.495\linewidth}
\centering
\includegraphics[scale=0.185]{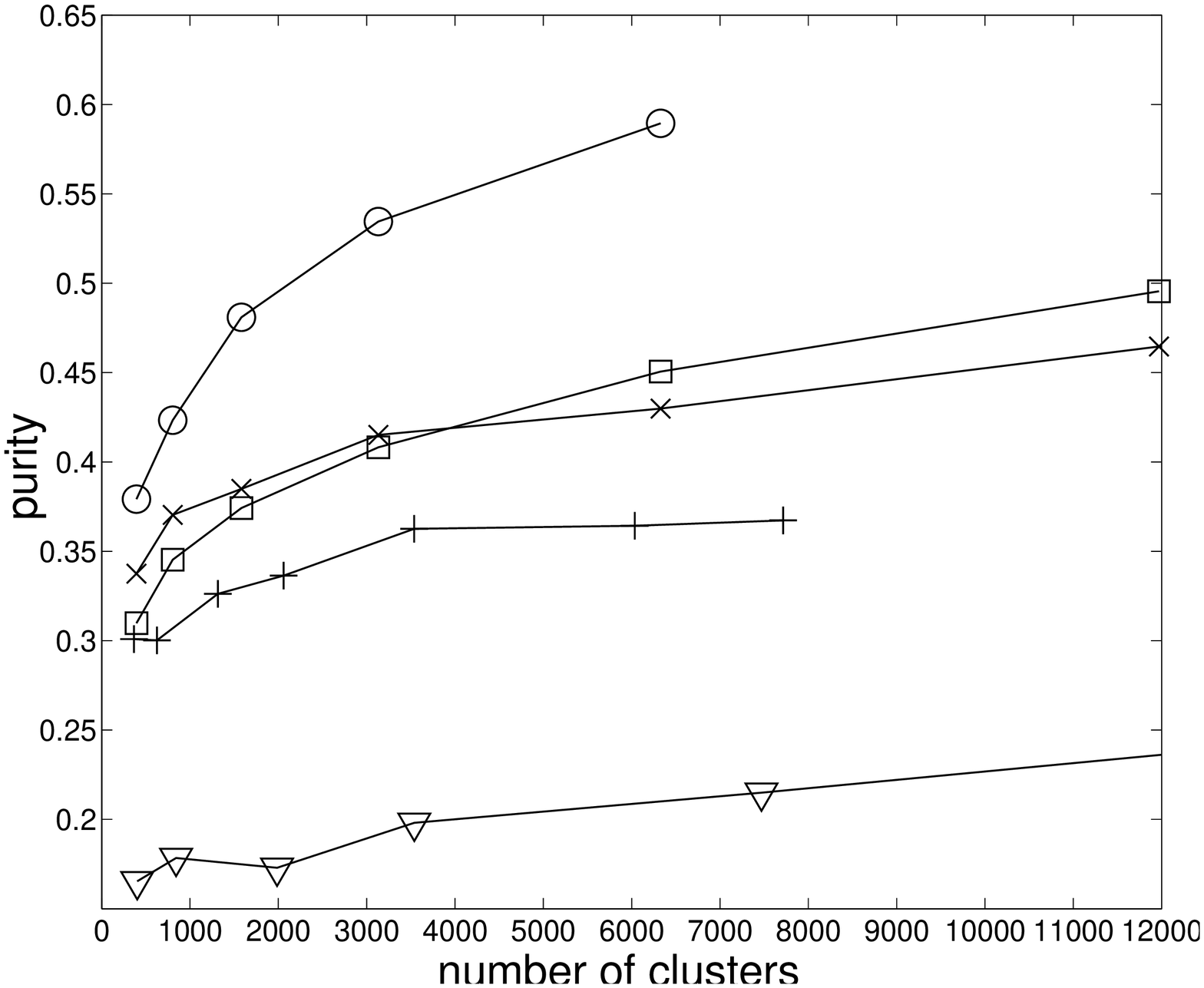}
\includegraphics[scale=0.185]{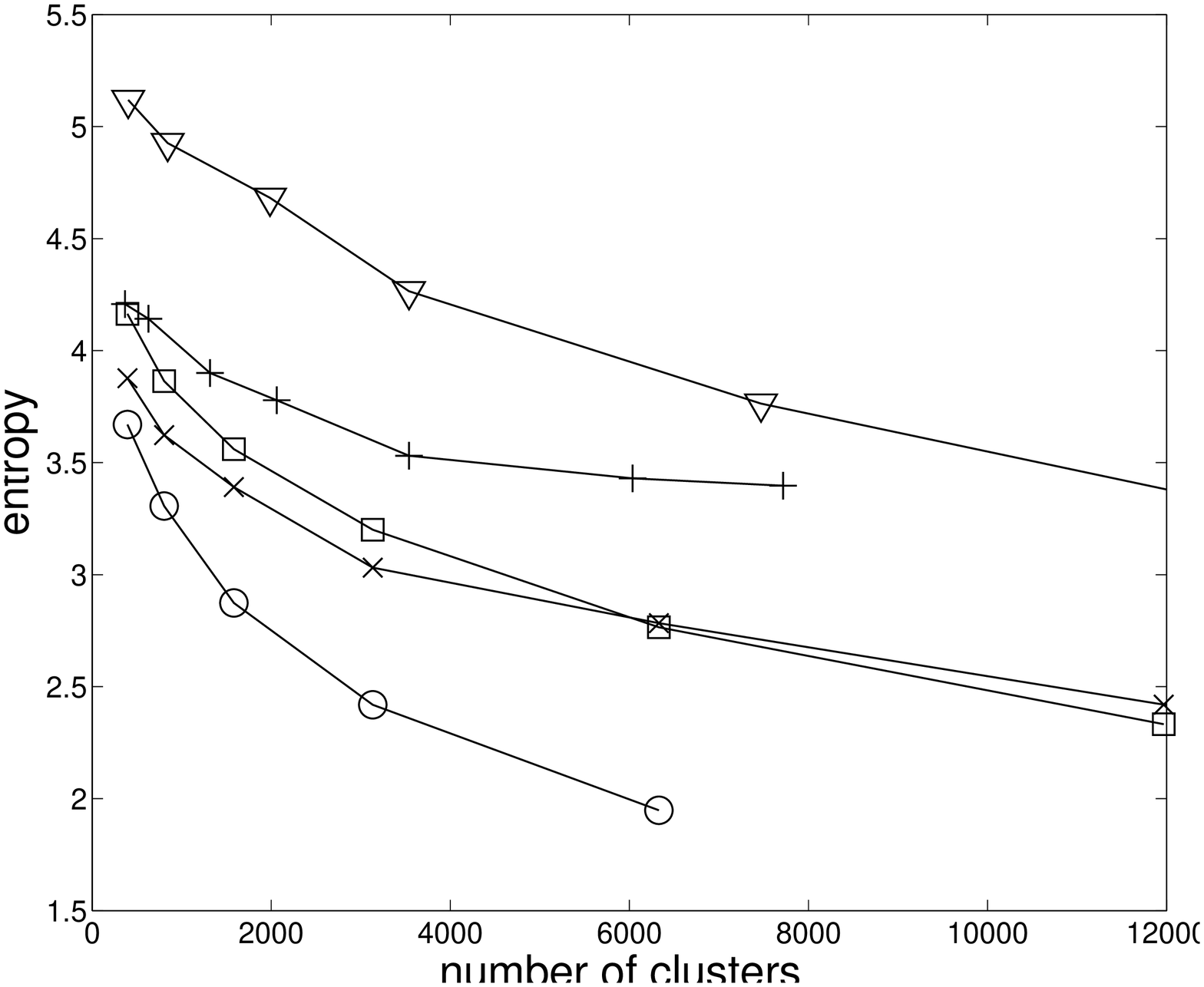}
\includegraphics[scale=0.185]{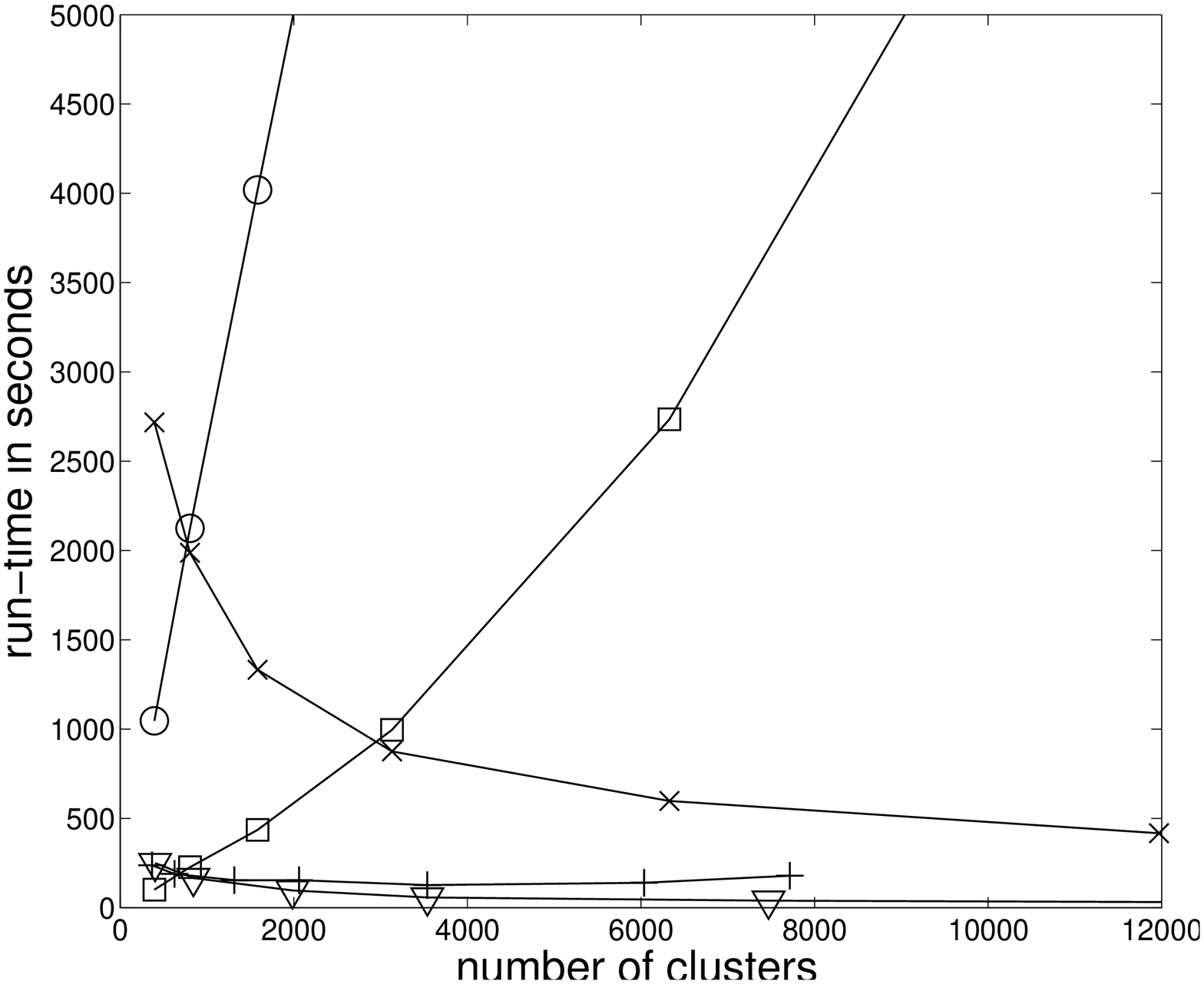}
\caption{RCV1}
\label{fig:rcv}
\end{minipage}
\end{figure}

K-tree produces large numbers of clusters in significantly less time. The graphs show an early poor run-time performance of K-tree. This has been artificially inflated by choosing a large tree order. One can choose a smaller tree order and find a smaller number of clusters higher in the tree. The computational cost of k-means dominates when choosing a large K-tree order. Note that the k-means algorithms in K-tree and CLUTO differ. K-tree runs k-means to convergence using dense vectors. CLUTO stops after a specified number of iterations and uses sparse vectors. Using the medoid K-tree further increases run-time performance and the combination of both K-tree variants falls somewhere in between. These solutions provide significant performance gains when large numbers of clusters are required. This leaves a trade off between performance and quality. Figures~\ref{fig:inex} and \ref{fig:rcv} show quality relationships and run-time performance.

\section{Applications in Information Retrieval}
The performance gain of K-tree is most pronounced when many small clusters are required. Many practical applications require this. For example, many clusters are useful for collection selection when deciding how to spread a collection across many machines. When using clustering for link discovery, a large number of small clusters are useful. A small number of highly semantically related documents are candidates for linking.

Due to the dynamic nature of the algorithm, clusters can be produced incrementally without having to process all documents at once. This also allows for easy updates as new documents arrive.

The tree structure can be used for interactive collection exploration. A user can start browsing from any point in the tree and generalise or specialise what they are viewing by traversing up or down the tree. A list of documents ranked against the current cluster centre can be displayed. This allows the user to conceptualise clusters as they move through the tree.

\bibliographystyle{plain}

\end{document}